\pdfoutput=1

\documentclass[sigconf]{acmart}
\usepackage{hyperref}
\usepackage{booktabs} 
\usepackage[outdir=./]{epstopdf}
\usepackage{graphicx}
\usepackage{array, multirow}

\def \f {\mathbf{f}}

\def \x {\mathbf{x}}
\def \u {\mathbf{u}}

\def \u {\mathbf{u}}

\def \I {\mathbb{I}}
\def \R {\mathbb{R}}
\def \W {\mathbf{W}}

\def \bq {\begin{eqnarray}}
\def \eq {\end{eqnarray}}
\def \bqs {\begin{eqnarray*}}
\def \eqs {\end{eqnarray*}}

\setcopyright{rightsretained}

\acmDOI{10.475/123_4}

\acmISBN{123-4567-24-567/08/06}

\acmYear{2018}
\copyrightyear{2018}

\begin{document}
\title{URLNet: Learning a URL Representation with Deep Learning for Malicious URL Detection}

\author{Hung Le, Quang Pham, Doyen Sahoo, Steven C.H. Hoi}
\affiliation{%
  \institution{School of Information Systems, Singapore Management University\\\{hungle, hqpham.2017, doyens, chhoi\}@smu.edu.sg}
}

\begin{abstract}
	Malicious URLs host unsolicited content and are used to perpetrate cybercrimes. It is imperative to detect them in a timely manner. Traditionally, this is done through the usage of blacklists, which cannot be exhaustive, and cannot detect newly generated malicious URLs. To address this, recent years have witnessed several efforts to perform Malicious URL Detection using Machine Learning. The most popular and scalable approaches use lexical properties of the URL string by extracting Bag-of-words like features, followed by applying machine learning models such as SVMs. There are also other features designed by experts to improve the prediction performance of the model. These approaches suffer from several limitations: (i) Inability to effectively capture semantic meaning and sequential patterns in URL strings; (ii) Requiring substantial manual feature engineering; and (iii) Inability to handle unseen features and generalize to test data. To address these challenges, we propose URLNet, an end-to-end deep learning framework to learn a nonlinear URL embedding for Malicious URL Detection directly from the URL. Specifically, we apply Convolutional Neural Networks to both characters and words of the URL String to learn the URL embedding in a jointly optimized framework. This approach allows the model to capture several types of semantic information, which was not possible by the existing models. We also propose advanced word-embeddings to solve the problem of too many rare words observed in this task. We conduct extensive experiments on a large-scale dataset and show a significant performance gain over existing methods. We also conduct ablation studies to evaluate the performance of various components of URLNet.
\end{abstract}


\begin{CCSXML}
	<ccs2012>
	<concept>
	<concept_id>10002978.10002997.10003000.10011612</concept_id>
	<concept_desc>Security and privacy~Phishing</concept_desc>
	<concept_significance>500</concept_significance>
	</concept>
	<concept>
	<concept_id>10010147.10010257.10010293.10010294</concept_id>
	<concept_desc>Computing methodologies~Neural networks</concept_desc>
	<concept_significance>500</concept_significance>
	</concept>
	</ccs2012>
\end{CCSXML}

\ccsdesc[500]{Security and privacy~Phishing}
\ccsdesc[500]{Computing methodologies~Neural networks}

\keywords{URLNet, Malicious URL Detection, Deep Learning}

\maketitle

\section{Introduction}
	Malicious URLs are one of the primary mechanisms to perpetrate cyber crimes. They host unsolicited content and attack unsuspecting users, making them victims of various types of scams (theft of money, identity theft, malware installation, etc.). This has resulted in billions of dollars worth of losses every year \cite{hong2012state}. It has thus become imperative to design robust techniques to detect malicious URLs in a timely manner. Traditionally, and most popularly, this detection is done through the usage of blacklisting methods. These are essentially lists of URLs collected by anti-virus groups which are "known" to be malicious. They are often collected through crowd sourcing solutions (e.g. PhishTank \cite{opendns2016phishtank}). While these methods are fast (requiring a simple database lookup), and are expected to have low False Positive rates, a major shortcoming is that they cannot be completely exhaustive, and in particular they fail against newly generated URLs. This is a severe limitation as new URLs are generated everyday. To address these limitations, there have been several attempts to solve this problem through the use of machine learning \cite{sahoo2017malicious}. In particular machine learning models offer the ability to generalize their predictions on new unseen URLs. 
	
	Malicious URL Detection through machine learning typically comprises two steps: first to obtain an appropriate feature representation from the URL, and second, to use this representation of the URL to train machine learning based prediction models. The first step of obtaining feature representation deals with obtaining useful information about the URL that can be stored in a vector so that machine learning models can be applied to it. Various types of feature have been considered, including lexical features, host-based features, content features, and even context and popularity features \cite{sahoo2017malicious}. However, the most commonly used features are lexical features, as they have demonstrated good performance and are relatively easy to obtain \cite{ma2009beyond,blum2010lexical}. Lexical features describe the lexical properties obtained from the URL string. These include statistical properties such as length of the URL, number of dots, etc. In addition, Bag-of-Words like features are often used. Bag-of-Words indicate whether a particular word or string appears in the URL or not. Consequently, every unique word in the training dataset becomes a feature. Using these features, in the second step, prediction models such as SVMs are trained. These models can be viewed as a form of fuzzy blacklists. 
	
	While the above approaches have shown successful performance, they suffer from several limitations, particularly in the context of very large scale Malicious URL Detection: 
	(i) \emph{Inability to effectively capture semantic or sequential patterns}: Existing approaches rely on using Bag-of-Words features, which essentially give information about the presence of a word in the URL. They fail to effectively capture the sequence in which words (or characters) appear in the URL String; 
	(ii) \emph{Require substantial manual feature engineering}: many of these approaches require expert guidance to determine the important features for the task (e.g. which statistical properties of the URL to use, what type of n-gram features would be better, etc.);
	(iii) \emph{Inability to handle unseen features}: During prediction, test URLs are likely to contain new words that did not exist in the training data. Under these circumstances, the trained models are unable to extract any useful information about the URL from these words. Moreover, the number of unique words in URLs can be extremely large, causing severe memory constraints while training models.
	
	To address the above issues we propose URLNet, a Deep Learning based solution for Malicious URL Detection. Deep Learning \cite{lecun2015deep,schmidhuber2015deep,goodfellow2016deep} uses layers of stacked nonlinear projections in order to learn representations of multiple levels of abstraction. It has demonstrated state of the art performance in many applications (computer vision, speech recognition, natural language processing, etc.). In particular, Convolutional Neural Networks (CNNs) have shown promising performance for text classification in recent years \cite{kim2014convolutional,zhang2015character}. Following their success, we propose to use CNNs to learn a URL embedding for Malicious URL Detection. 
	
	Specifically, URLNet receives a URL string as input and applies CNNs to both characters and words in the URL.  
	For Character-level CNNs we first identify unique characters in the training corpus, and represent each character as a vector.  Using this, the entire URL (a sequence of characters) is converted to a matrix representation, on which convolution can be applied. Character CNNs  identify important information from certain groups of characters appearing together which could be indicative of maliciousness. 
	For Word-level CNNs, we first identify unique words in the training corpus, delimited by special characters. Using a word-embedding matrix, we obtain a matrix representation of the URL (which in this context, is a sequence of words). Following this convolution can be applied. Word-level CNNs identify useful patterns from certain groups of words appearing together. However, using word-embeddings faces some challenges: (i) it cannot obtain embeddings for new words at test time; and (ii) too many unique words (specifically in malicious URL detection) - resulting in memory constraints while learning word embeddings. To alleviate these, we propose advanced word-embeddings where the word embedding is learned using the character-level information of each word. This also helps recognize subword level information. Both Character-level and Word-level CNNs are jointly optimized to learn the URLNet prediction model.
	
	URLNet allows us to alleviate the shortcomings of traditional approaches such that (i) Character and Word CNNs automatically identify and learn the semantic and sequential patterns in which the characters and words appear in the URL; (ii) Expert feature engineering required is reduced, since the CNN automatically learns features to represent the URL, and we do not rely on any other complex or expert features for the learning task; and (iii) The model learns patterns based on both character and word embeddings. Due to the limited number of characters, this character embedding can generalize to new URLs easily. For word-embeddings, even if the test URLs contain new unseen words, the character-based (advanced word) embedding of the words still allows us to obtain representation for these new words. This way URLNet has superior generalization ability compared to existing approaches. We conduct extensive experiments, analysis and ablation studies to show the efficacy of proposed method.

\section{Malicious URL Detection}
	\subsection{Problem Setting}
	Our goal is to classify a given URL as malicious or not. We do this by formulating the problem as a binary classification task. Consider a set of $T$ URLs, $\{(\u_1,y_1), \dots, (\u_T, y_T)\}$, where $\u_t$ for $t=1,\dots,T$ represents a URL, and $y_t \in \{-1,+1\}$ denotes the label of the URL, with $y=+1$ being a malicious URL, and $y_t = -1$ being a benign URL. 
	The first step in the classification procedure is to obtain a feature representation $\u_t \rightarrow \x_t$ where $\x_t \in \R^n$ is the $n-$dimensional feature vector representing URL $\u_t$. The next step is to learn a prediction function $\f: \R^n \rightarrow \R$ which is the score predicting the class assignment for a URL instance $\x$. The prediction made by the function is denoted as $\hat{y_t} = \text{sign}(\f(\x_t))$. The aim is to learn a function that can minimize the total number of mistakes ($\sum_{t=1}^T \I_{\hat{y_t} \ne y_t}$) in the entire dataset. This is often achieved by minimizing a loss function. Many types of loss functions can be used, which may also include a regularizer term.
	For Deep Learning, the function $\f$ is represented as a Deep Neural Network, such as a Convolutional Neural Network.
	
	\subsection{Lexical Features}
	\label{sec:lexical}
	Lexical features have often been adopted for the first stage in which the raw URL $\u$ is converted to a feature vector $\x$. A URL is split into component words which are delimited by special characters in the URL. Using the entire training corpus, all unique words in the training dataset are identified to construct a dictionary, and each word $w_i$ becomes a feature. Given $M$ distinct features, each URL $\u_t$ is then mapped to a vector $\x_t \in \R^M$, such that $i^{th}$ element in $\x_t$ is set as $1$ if word $w_i$ is present in the URL, and $0$ otherwise.  In addition to these Bag-of-Words features, other statistical features are commonly used, including the length of URL, lengths of different segments in the URL, number of dots, etc.	
	
	Since the number of unique words can be quite large, the feature size of this dataset is also proportionally large, typically more than the number of URLs present in the corresponding training dataset. Additionally, to retrain the model with new data, the feature size grows, and so does the model size. Here, we describe how exactly using these features are responsible for limitations of existing methods. 
	Three major limitations of these features are: (i) Lack of information about the sequence in which the characters or words appear in the URL. Some strategies have been used to exploit sequential information by creating a separate dictionary for every segment of the URL \cite{ma2009beyond,ma2009identifying,blum2010lexical}. For example, this can help distinguish between "com" appearing in the Top Level Domain and "com" appearing the Path of URL. Even then, such strategies do not account for the sequence in which the words or characters appear within a specific segment of the URL. Moreover, these methods cannot exploit information from substrings that may appear within a word of the URL; (ii) Inability to obtain information from rare words. Many words in the URL training corpus appear only once, and if training models such as SVMs are used, these features are unable to provide any useful information; and (iii) Inability to interpret new words in test URLs. Since the new words have never appeared in the training data, the models fail to extract useful information about the URLs from these words.  
	
	To address these issues, we propose URLNet. 

\section{URL Net}
\begin{figure*}[htbp]
		\centering
		\includegraphics[width=0.99\textwidth]{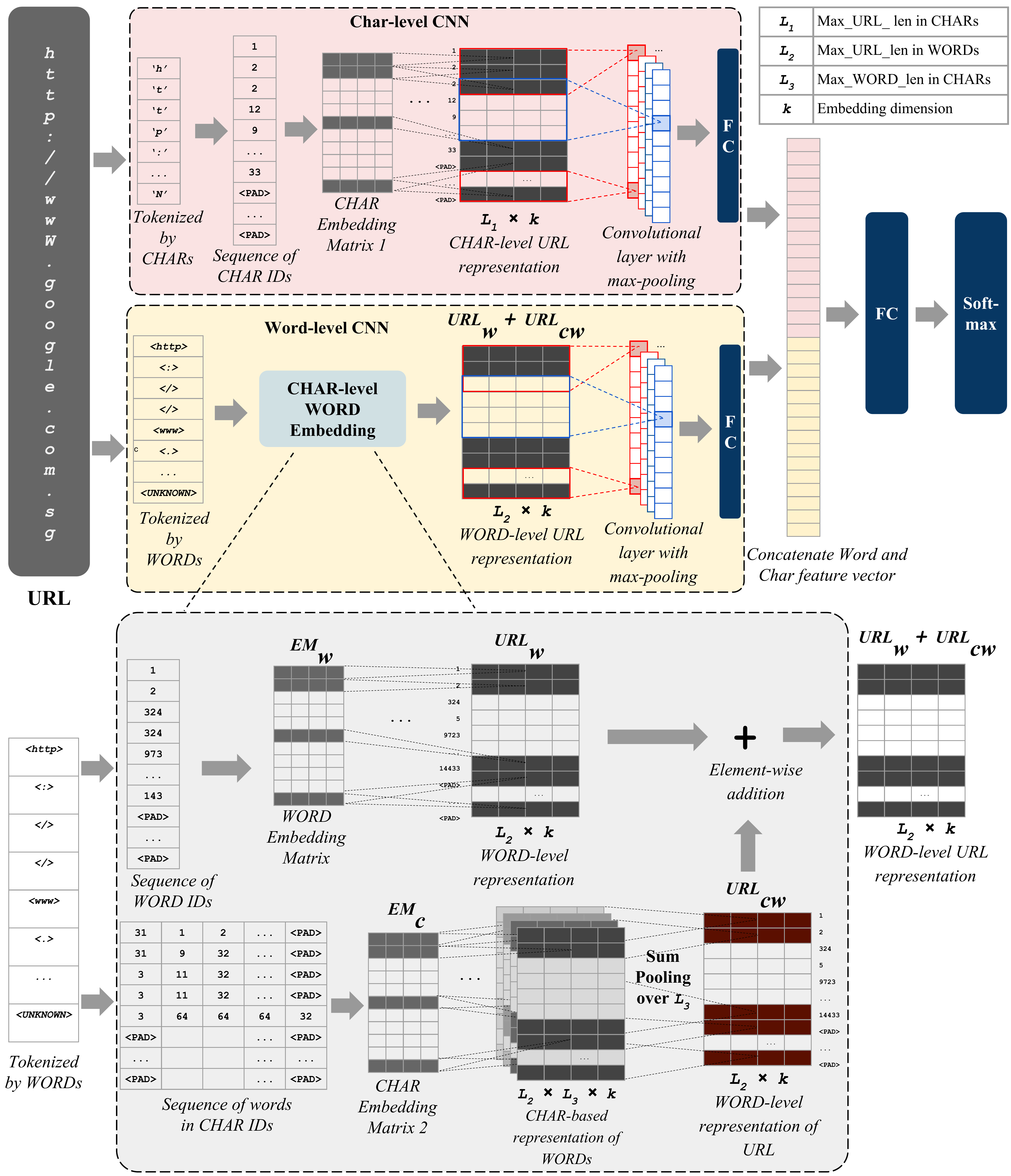}
		\caption{URLNet - Deep Learning for Malicious URL Detection. It comprises two branches with CNNs. The first branch is a Character-level CNN where the character embedding is used to represent the URL. The second is the Word-level CNN where a word-level embedding is used to represent the URL. The word embedding itself is a combination of the individual word's embedding and the character-level embedding of that word.}
		\label{fig:URLNET}
	\end{figure*}

	In this section, we describe the proposed Deep Learning based URLNet for Malicious URL Detection. The entire network can be visualized in Figure \ref{fig:URLNET}. 
	
	\subsection{Deep Learning for Malicious URL Detection}
	
	The underlying deep neural network for URLNet is a Convolutional Neural Network (CNN). CNNs have achieved extraordinary success in Computer Vision tasks \cite{krizhevsky2012imagenet,he2015deep}. Their designs allowed them to automatically learn the salient features in the images from raw pixel values. Eventually their principles were adopted for Natural Language Processing \cite{dos2014deep,johnson2014effective,kim2014convolutional,zhang2015character}, where the CNNs could learn useful structural information in text from raw values of word or character embeddings. In URLNet, the CNNs are used to learn structural information about the URL. Specifically CNNs are applied at both the character-level and word-level. Next, we describe a simple CNN for URL classification. 
	
	A URL $\u$ is essentially a sequence of characters or words (delimited by special characters). 
	We aim to obtain its matrix representation $\u \rightarrow \x \in \R^{L\times k}$, such that the instance $\x$ comprises a set of contiguous components $x_i, i = 1,\dots,L$ in a sequence, where the component can be a character or a word of the URL. Each such component is represented by an embedding such that $x_i \in \R^k$, is a k-dimensional vector. 
	
	Usually, this $k-$dimensional representation for a component is an embedding vector extracted from an embedding matrix that is randomly initialized and jointly learned with the rest of the model. In our work, we randomly initialize the embedding matrix, and learn it in the end-to-end optimization. 
	With this notation, an instance with a sequence of L components can be represented as:
	\bq
	\nonumber
	\x = \x_{1:L} = x_1 \oplus x_2 \oplus \ldots \oplus x_L
	\eq
	where $\oplus$ denotes the concatenation operator. 
	For the purpose of parallelization, usually all sequences are padded or truncated to the same length $L$.
	
	A CNN would convolve over this instance $\x \in \R^{L\times k}$ using a convolutional operator.	A convolution operation $\otimes$ of length $h$ consists of convolving a filter $\W \in \R^{k\times h}$ followed by a non-linear activation $f$ to produce a new feature:
	\bq
	\nonumber
	c_i = f(\W\otimes \mathbf{x}_{i:i+h-1} + b_i)
	\eq
	where $b_i$ is the bias.
	This convolution layer's output applies a filter $\W$ with a nonlinear activation to every $h-$length segment of its input, each of which is separated by a pre-defined stride value. These outputs are then concatenated to produce output $\mathbf{c}$  such that:
	\bq
	\nonumber
	\mathbf{c} = [c_1, c_2, \ldots, c_{L-h+1}]
	\eq
	After the convolution, a pooling step (either max or average pooling) is applied to reduce the feature dimension and to identify the most important features. 
	
	By using a filter $\W$ to convolve on every segment of length  $h$, the CNN is able to exploit the temporal relationship of length $h$ in its input.
	A CNN model typically consists of multiple sets of filters with different lengths ($h$), and each set consists of multiple filters. These are hyperparameters of the model that need to be set by the user. 
	A convolution followed by a pooling layer comprises a block in this deep neural network. There can be multiple such blocks that can be stacked on top of each other. 
	The pooled features from the final block are concatenated and passed to fully connected layers for the purpose of classification.
	The network can then be trained by stochastic gradient descent using backpropagation.
	
	URLNet uses multiple CNNs: one for character-level and one for word-level. Next we describe each component of URLNet in detail.

	\subsection{Character-level CNN for Malicious URL Detection}
	Here we present the key ideas for building Character level CNNs for Malicious URL Detection. We aim to learn an embedding that captures the properties about the sequence in which the  characters appear in a URL. 
	To do this, we first identify all the unique alphanumeric and special characters in the dataset. 
	Characters which appear less frequently (e.g. less than 100 times in a corpus of millions of URLs) are replaced with the unknown token denoted by <UNK>.
	We obtained $M=96$ unique characters including the <UNK> and <PAD> tokens. 
	We set the length of the sequence $L_1=200$ characters. URLs longer than 200 characters would get truncated from the $200^{th}$ character, and any URLs shorter than $200$ would get padded with the <PAD> token till their lengths reached $200$.
	
	Each character is embedded into a $k-$dimensional vector. In our work, we choose $k=32$ for characters. This embedding is randomly initialized and is learnt during training. For ease of implementation, these representations are stored in an embedding matrix $EM \in \R^{M \times k}$, where each row is the vector representation of a character. Using this embedding, each URL $\u$ is transformed into a matrix, $\u \rightarrow \x \in \R^{L_1\times k}$, where $k=32$ and $L_1=200$. 
	
	Using the URL matrix (for all the URLs $\x_t \forall t = 1, \dots, T$) as the training data, we can now add convolutional layers. We use $4$ types of Convolutional filters $\W \in \R^{k\times h}$, with $h = 3,4,5,6$ respectively. Thus, temporal patterns in a sequence of characters of lengths $3,4,5,6$ are learnt. 
	For each filter size, we use $256$ filters. This is followed by a Max-Pooling layer which is followed by a fully connected layer regularized by dropout. The result is concatenated with other branches of the URLNet, finally leading to the output layer. 
	
	Apart from the ability to learn structural patterns in the URL String, Character-level CNNs also allow for easily obtaining an embedding for new URLs in the test data, thus not suffering from inability to extract patterns from unseen words (like existing approaches). 
	As the total number of characters is fixed, the model size of the Character-level CNN remains fixed (unlike models based on words - where model size increases with data size). 
	However, Character-level CNN is not able to exploit information from long sequences of components in the URL. It also ignores word boundaries, making it difficult to distinguish special tokens in the data. Further, in scenarios where the malicious URLs try to mimic benign URLs by having a minor modification to one or few words of the URL \cite{chu2013protect}, Character-level CNN may struggle to identify this information. This is because a sequence of characters with similar spellings is likely to obtain a similar output from the convolutional filters. Thus, Character-level CNNs alone are not sufficient to comprehensively obtain structural information from the URL String, and it is necessary to consider word-level information as well. 
	
	\subsection{Word-level CNN for Malicious URL Detection}
	
	Word-level CNNs are similar to Character-level CNNs, except the convolutional operators are applied over words. We present a basic word-level CNN, followed by two advanced methods. 
	
		\subsubsection{Word-level CNNs}
			We first identify all the unique words that appear in the training corpus of the URLs. 
			Unlike Character CNNs where the number of unique characters is small and (usually) fixed, the number of unique words depends on the size of the training corpus, as new words can appear in every URL. We identify the unique words using the approaches in Section \ref{sec:lexical}, and \cite{ma2009beyond,blum2010lexical}. All unique words are obtained as a sequence of alphanumeric characters (including '-' and '\_'), separated by special characters (e.g. '.', '/', etc.). 
			We also use the <PAD> token as an additional word to make the lengths of the URLs uniform in terms of number of words ($L_2=200$).
			This set of unique words forms a dictionary for the URL training corpus. 
			
			In the next step, we obtain the $k-$dimensional vector representation for each word. In our work, we set $k = 32$, i.e., each word is embedded into a 32-dimensional vector. For $M$ unique words, we have to learn an embedding matrix $EM \in \R^{M\times k}$.
			Using this representation, all the URLs are converted to their respective matrix representation ($L_2 \times k$), on which the CNN is applied. We use the same CNN architecture as in Character CNNs, i.e., we use  $4$ types of Convolutional filters $\W \in \R^{k\times h}$, with $h = 3,4,5,6$ and for each filter size, we use $256$ filters. Here, the aim is to learn temporal properties from a sequence of words of length $3,4,5,6$ appearing together. This is followed by a Max-Pooling layer and the fully connected layer regularized by dropout. This output is then concatenated with the other branches of the URLNet.
			
			As the number of words can be extremely large (See Section \ref{sec:featureExtraction} for an example), the corresponding Embedding Matrix would become very large, and the model would suffer from memory constraints. As a result, all words that appeared only once in the entire training corpus (also called rare words) were replaced with a single <UNK> token. This would substantially reduce the memory constraints faced by word-based models. During test time, a lot of URLs would contain words that have not been seen during training. Therefore, during conversion of a test URL to matrix form, the unseen words are also replaced with an unknown token <UNK>.   
			
		\subsubsection{Special Characters as Words}
			\label{sec:special}
			While modeling Word-level CNNs, the words are obtained as a series of alphanumeric characters separated by special characters. This approach does not consider using useful information that can be obtained by evaluating the special characters in the URL String. There are two types of information missed out on: (i) The distribution and types of special characters used; and (ii) The temporal relation of the appearance of words around special characters. To alleviate these issues, we propose to use special characters in the URL Strings as unique words. Conventional approaches in Deep Learning for NLP rarely exploit special characters as words (apart from direct usage in Character-level CNNs). However, we hypothesize that special characters offer significant information gain for Malicious URL Detection because special characters are more frequent and relevant in the context of URLs than normal natural languages. As URL does not follow normal semantic syntax, special characters can play an important feature and should be considered with words.
			
		\subsubsection{Improved Word Embedding Using Character-level Word Embedding}
			The above model does not use the rare words due to memory constraints. Further, it is not able to obtain an effective embedding for new words in test URLs. To address these concerns, we propose to obtain a character-level embedding for each word. In contrast to the previous scenario where the word embedding is obtained directly from the word embedding matrix (which is learnt during training), we obtain the word embedding as a combination of the original word embedding and the embeddings of the individual characters  in that word. 
			
			Specifically, we maintain 2 Embedding Matrices: one for words ($ EM_w \in \R^{L_2\times k}$), and one for characters ($EM_c \in \R^{L_1\times k}$). Note that the character embedding matrix here is different from the character embedding matrix used in Character-level CNN. While the Character Embedding Matrix for the Character-level CNN aims to learn character representation based on the full URL, $EM_c$ is more localized, and aims to learn the appropriate character embedding to effectively represent the words.
			While obtaining the URL Matrix representation, we first get $URL_w \in \R^{L_2 \times k}$ representation based on $EM_w$. 
			Next, we obtain $L_3 \times k$ matrix representation of each word in a URL (using $EM_c$), where each word (as obtained in Section \ref{sec:special}) is padded or truncated to be a sequence of $L_3 = 20$ characters. This matrix is summed up to obtain a $1 \times k$ vector embedding for the word. This is applied to all $L_2$ words in the URL to give us the URL matrix representation $URL_{cw} \in \R ^ {L_2 \times k}$, which is termed as character-level word embedding of the URL. The final URL matrix representation is simply the sum of these two matrices $URL_w + URL_{cw}$. This entire approach is visualized in URLNet in Figure \ref{fig:URLNET}.
			
			During training, the words that appeared only once (rare words) in the entire training data were ignored and converted to a single <UNK> token. However, for each of these words, a character-level embedding was used, thus giving each of the rare words a (mostly) unique representation even during training. 
			In a similar fashion, during test time we are able to obtain a unique word embedding even for the new words not used in training.
			Thus, the proposed character-level word embedding addresses the issues of memory constraints of word-based models, is able to exploit information from rare words, and also able to obtain a richer representation to capture subword level information.
			
	\subsection{Model Configuration}
\begin{figure}[h]
	\centering
	\includegraphics[width=0.45\textwidth]{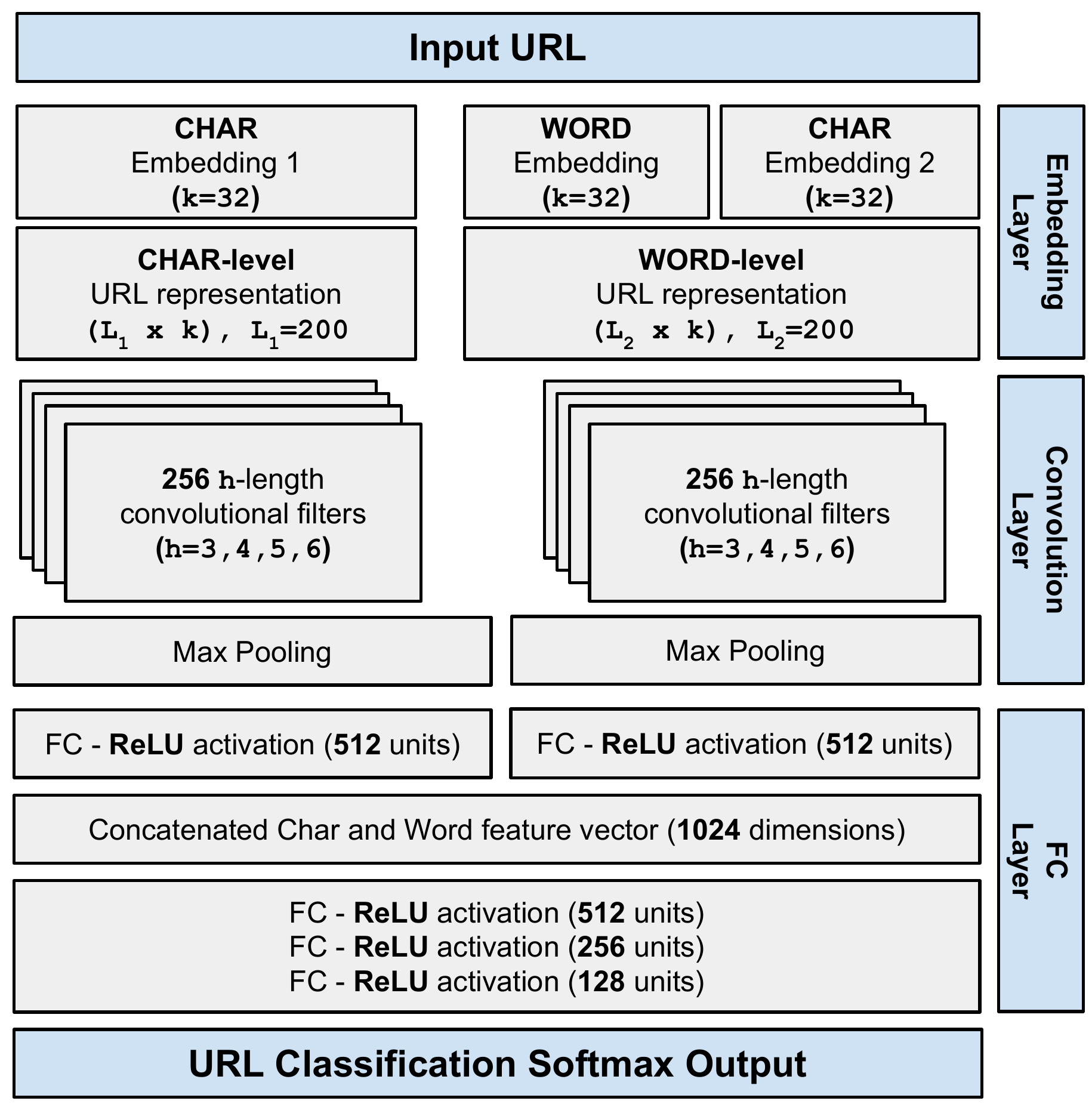}
	\vspace{-0.1in}
	\caption{Configuration of URLNet}
	\label{fig:model}
	\vspace{-0.2in}
\end{figure}

	Finally, we present the details of the URLNet model configuration.
	The overview can be seen in Figure \ref{fig:model}. The raw URL input string is processed by 2 branches: a character-level branch and a word-level branch. The character-level branch gives a character-level representation of the URL, while the word-level branch does the same with words.	
	The word-level branch itself is split into word-embedding and character-level word embedding, which are combined to finally give the word-level URL representation. Convolutional operations are applied to both these branches, followed by a fully connected (FC) layer, which is regularized by dropout for both the branches. Then the outputs are concatenated. This is followed by 4 fully-connected layers finally leading to the output classifier. This model is then trained by an optimizer using Backpropagation.

\section{Experiments}
	\subsection{Large Scale Dataset}	
	\subsubsection{Dataset Collection}
	We collected a large corpus of labeled URLs from VirusTotal \footnote{https://www.virustotal.com/}. VirusTotal is an antivirus group whose services are often used to validate whether a given query URL is malicious or not. Given an input URL, VirusTotal scans through 64 different blacklists (e.g.  CyberCrime, FraudSense, BitDefender, Google Safebrowsing, etc.), and reports how many of these blacklists contain the input URL. If none of the blacklists contain the given URL, it is assumed that the URL is benign. The higher the number of blacklists the URL is detected by, the higher is our confidence that the given URL is indeed malicious.
	
	We crawled all URLs queried in VirusTotal in the period between May, 2017 to June, 2017, to build our training datatset. From the resulting URLs, we removed any duplicates. We remove the duplicates in order to examine the generalization performance of the models on unseen URLs. However, this makes our benchmark more challenging that traditional (possibly real world) settings. We observed that there were a few dominant domains which occurred very frequently. To reduce any bias, we limited the frequency of any URL domain to be less than 5\%. All URLs that did not appear in any blacklist were labelled Benign, and all URLs that appeared in 5 or more blacklists were labelled Malicious. All URLs that appeared in 1, 2, 3, or 4 blacklists were discarded, due to lack of certainty about their true labels. This gave us millions of URLs with roughly 94\% Benign and 6\% Malicious. 
	
	The resulting URLs were sorted according to the timestamp at which they were queried. After sorting, from the first 60\% of the URLs, we randomly selected 5 million URLs for training, and from the last 40\%, we randomly selected 10 million URLs for testing.  
	The sorting was done to avoid any "look-ahead" bias while training the models. Other details about this corpus can be seen in Table \ref{tab:largeData}.
	
	\begin{table}[h]		
		\caption{URL Dataset crawled from VirusTotal}
		\begin{tabular}{|c|ccc|}
			\hline
			& \textbf{Benign URLs} & \textbf{Malicious URLs} & \textbf{Total} \\ \hline
			\textbf{Training } &      4,683,425     &        316,575       &   5,000,000   \\
			\textbf{Testing }  &      9,366,850     &        633,150      &   10,000,000   \\
			\textbf{Total}   &      14,050,275      &        949,725        &  15,000,000 \\ \hline
		\end{tabular}
		\label{tab:largeData}
	\end{table}

	\subsubsection{Feature Extraction}
	\label{sec:featureExtraction}
	
	Features are extracted from the raw URLs before training a model. For Character CNNs no feature extraction is required, as the CNN directly operates at the character level. For Word CNN, we extracted lexical features in the form of Bag of Words. For a given training dataset, all unique words are identified and a dictionary is constructed. The number of unique words is denoted by $M$, and this value varies with the size of the training dataset. This set of features are called \emph{Whole URL BoW}.		

	We also consider other types of features extracted by security experts for baseline models along the lines of the work in \cite{ma2009identifying,blum2010lexical,lin2013malicious}:
		\begin{itemize}
			\item URL Component Tokenization (UCT): In \cite{ma2009identifying}, URL is divided into primary domain, path, last path token, and top level domain - and BoW dictionary is constructed for each component. This helps capture some order in terms of sequential information in the URL.
			\item Position Sensitive \& Bigrams (PSB): In \cite{blum2010lexical}, in addition to UCT features, specific tokens such as \emph{Domain\{1\}} and \emph{Path\{2\}} are extracted to account for sequential information of tokens within the URL primary domain and the URL path. Here \emph{\{x\}} refers to the zero based token distance from the right most end of a URL part. In addition, token bigrams in primary domain and path are also extracted to further distinguish legitimate and malicious URLs. For example, the feature \emph{Domain\{1\}\{0\}} refers to the token bigram consisting of the second last token and the last token in the URL primary domain.  
			\item Character Trigrams: In \cite{lin2013malicious}, in addition to UCT features, three character length sliding window is used on the URL domain name to generate character trigram tokens. This is to tackle malicious URLs with subtly modified domain names. We also followed the same feature processing as \cite{lin2013malicious} by only extracting the argument names and discarding the argument values in the URL path. 
		\end{itemize}
	In addition to the BoW Features, as suggested by the above approaches, a variety of statistical properties of the URL were used, such as the length of the entire URL, the length of the hostname, and the number of dots in the URL. For the baseline \cite{lin2013malicious}, we also computed hand designed statistical features such as alphabet entropy measure, character continuity rate, and number rate. We call these statistical features as "Expert Features", as these are hand-designed by experts. 

	Further details on these Lexical Features can be seen in Table \ref{tab:features}. We show the number of words or features for a corpus of 1 million and 5 million URLs. Basic BoW refers to the features obtained following traditional BoW approaches. Advanced BoW refers to features such as bigram and (character) trigram features. As can be seen the total number of lexical features can be very large, and keeps increasing with the data size. Correspondingly, the size of word-based models also keeps increasing.

 \begin{table*}[htbp]
	\centering
	\caption{Number of lexical features (words) and expert features and the variation of Feature Size with size of Training Data}
	\label{tab:features}
	\begin{tabular}{|l|p{1.5cm}p{1.5cm}p{1.5cm}|l|p{1.5cm}p{1.5cm}p{1.5cm}|l|}
		\hline
		& \multicolumn{4}{c|}{\textbf{Training Size = 1m}}                                              & \multicolumn{4}{c|}{\textbf{Training Size = 5m}}                                              \\ \hline
		& \textbf{\#Basic BoW} & \textbf{\#Advanced BoW} & \textbf{\#Expert Features} & \textbf{Total} & \textbf{\#Basic BoW} & \textbf{\#Advanced BoW} & \textbf{\#Expert Features} & \textbf{Total} \\ \hline
		\textbf{Whole URL BoW}                       & 1,372,417            & N/A                     & N/A                        & 1,372,417      & 5,531,997            & N/A                     & N/A                        & 5,531,997      \\ 
		\textbf{UCT \cite{ma2009identifying}}    & 2,015,183            & N/A                     & 3                          & 2,015,186      & 7,952,252            & N/A                     & 3                          & 7,952,255      \\ 
		\textbf{PSB \cite{blum2010lexical}} & 2,015,183            & 4,845,517               & 17                         & 6,860,717      & 7,952,252            & 16,666,144              & 17                         & 24,618,413     \\ 
		\textbf{Character Trigrams \cite{lin2013malicious}}                  & 1,242,340            & 54,368                  & 69                         & 1,296,777      & 4,953,031            & 58,154                  & 69                         & 5,011,254      \\ 
		\textbf{Combined}                            & 2,015,183            & 4,899,885               & 76                         & 6,915,144      & 7,952,252            & 16,724,298              & 76                         & 24,676,626     \\ \hline
	\end{tabular}
\end{table*}

	\subsubsection{Word-frequency Distribution}
\begin{figure}[h]
		\centering
		\includegraphics[width=.45\textwidth]{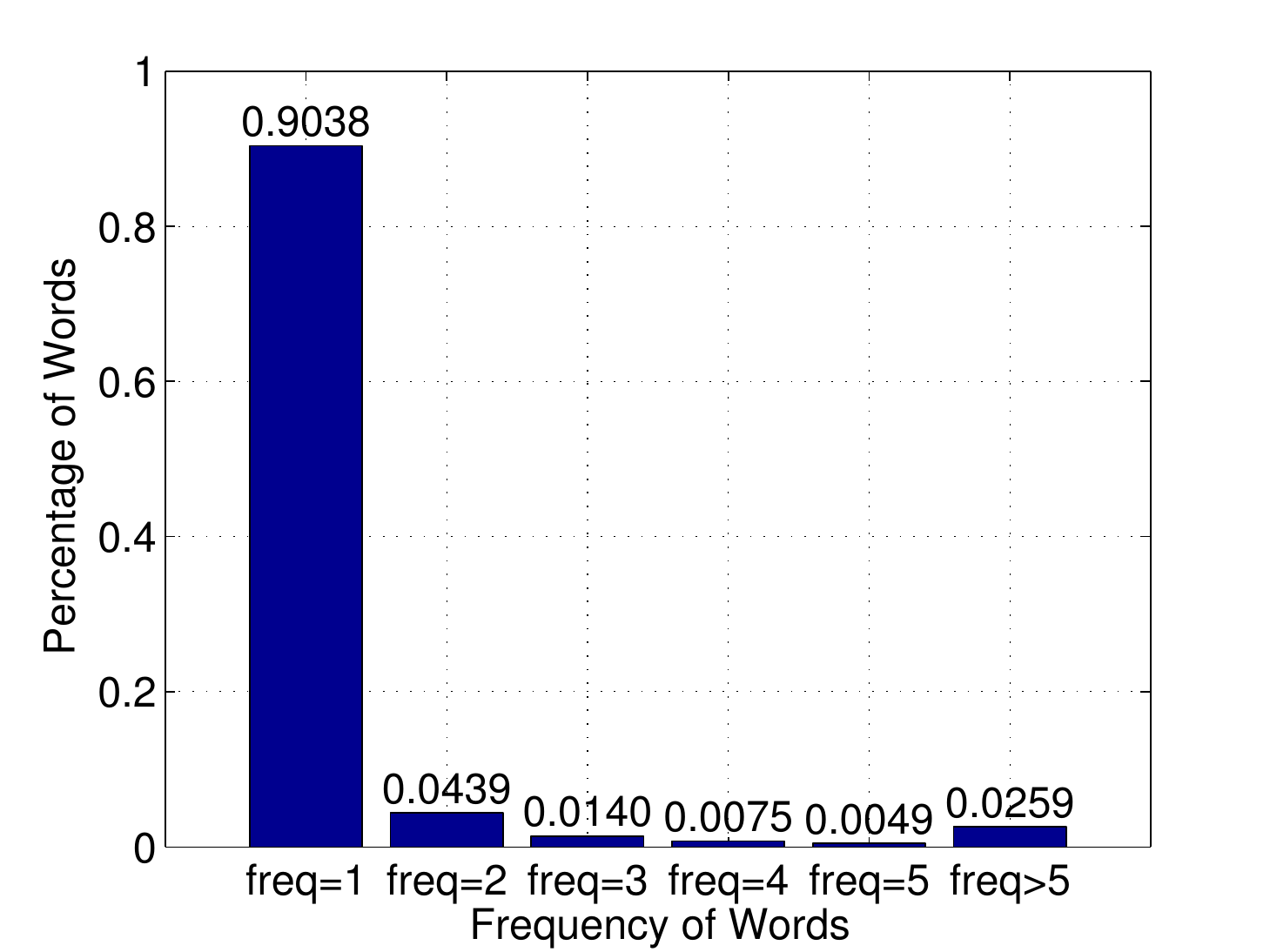}
		\caption{Word Frequency Distribution. More than 90\% of the words appear only once in the entire training corpus. Storing all the words, and learning their embeddings during training can be computationally prohibitive. }
		\label{fig:wordFrequency}
	\end{figure}

	Here, we look at the frequency distribution of the words in dataset. We obtain the whole URL BoW features for 1 million URLs, and plot the percentage of words based on their frequency of occurrence in the entire training data in Figure \ref{fig:wordFrequency}. As can be seen 90\% of the words in the training corpus appear only once. This implies that over 90\% of the unique words in a URL training corpus are very rare. While rare words make up a majority in the dictionary, storing them requires a lot of memory and is not feasible for large data sets. For example, our training data with 5 million URLs consists of over 5 million unique words. With an embedding size of $k=32$, these words require more than 150 million parameters for just the embedding matrix (similar to the number of parameters as a VGG-16 network\cite{simonyan2014very}). When scaling to even larger training datasets, storing this matrix becomes infeasible. By ignoring rare words, we are able to resolve the memory issue and make it possible to train word-related models on large data sets. Moreover, our proposed Character-level word embedding allows us to obtain representations of even the rare words (without storing the embedding for these words), and also captures local subword information for each word. 
	
	\subsection{Evaluation of URLNet}	
	
%	\begin{figure*}
%		\centering
%		\subfigure{
%			\includegraphics[width=.32\textwidth]{auc1.eps}}
%		\subfigure{
%			\includegraphics[width=.32\textwidth]{auc2.eps}}
%		\subfigure{
%			\includegraphics[width=.32\textwidth]{auc3.eps}}
%		\caption{ROC Characteristics of Different Methods. (a) ROC of baseline SVM, Character-level CNN, Simple Word-Level CNN and URLNet; (b) ROC analysis of different Word-level CNNs; (c) ROC analysis of different URL variants.}
%		\label{fig:weightDistribution}
%	\end{figure*}

		\subsubsection{Experimental Settings and Baselines}
		We constructed 2 training corpuses: first a set of 1 million URLs randomly sampled from the 5 million, and the second includes the entire 5 million URLs. 
		The sampling was done such that the proportion of malicious and benign URLs was still maintained. 
		
		As a baseline model, we used L1-regularized L2-loss SVM (implemented in Liblinear \cite{fan2008liblinear}). The SVM was trained on the 5 sets of baseline lexical features described in the previous subsection, according to  \cite{ma2009identifying,blum2010lexical,lin2013malicious}. The baseline methods include SVMs trained on: i) \textbf{Whole URL BoW}; (ii) \textbf{UCT} \cite{ma2009identifying}; (iii) \textbf{PSB}\cite{blum2010lexical}; (iv) \textbf{Character Trigrams} \cite{lin2013malicious}; and (v) \textbf{Combined}: a combination of all these features. We compared the baselines with the proposed URLNet: 
		\begin{itemize}
			\item \textbf{URLNet(Character-Level)} - only the Character-level CNN, 
			\item \textbf{URLNet(Simple Word-Level)} - only the Word-level CNN,
			\item \textbf{URLNet(Full)} - which is the end-to-end framework combining both character-level and word-level CNNs (including special characters and character-level word embedding).
		\end{itemize}
		For Word-level CNN, the words obtained were based on the Whole URL BoW (i.e. there was no separate dictionaries for different parts of the URL). All Deep Learning models were implemented using both Keras \cite{chollet2015keras} and Tensorflow \cite{abadi2016tensorflow}\footnote{https://github.com/Antimalweb/URLNet}, and optimized using the Adam Optimizer\cite{kingma2014adam}. A Dropout rate of $0.5$ was used in the convolutional layers. All the models were tested on the Test corpus of 10 million URLs and were evaluated on the basis of Area under the ROC Curve (AUC) (due to the imbalanced nature of the dataset). In addition, we also compare model performance in terms of True Positive Rates at different levels of False Positive Rate to observe the detection rate of malicious URLs at a given false positive rate (or false alarm rate).  
		
		\subsubsection{Results}
\begin{figure}[ht]
			\includegraphics[width=.4\textwidth]{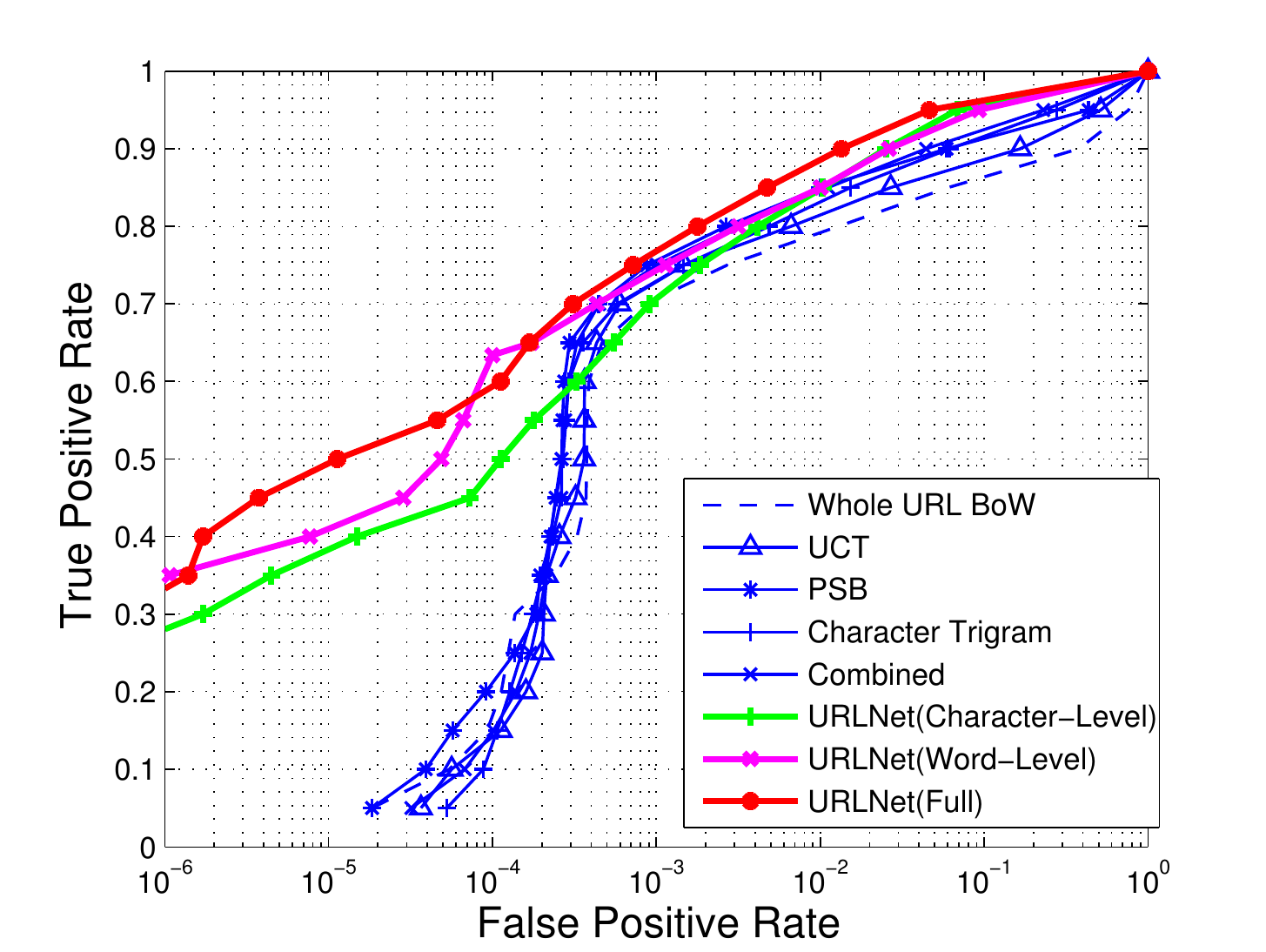}
			\caption{Area Under ROC Curve (Trained on 1m, Tested on 10m). URLNet(Full) is slightly worse than URLNet(Word-level) at FPR = $10^{-4}$, but better otherwise. URLNet(Full) is consistently better than URLNet(Character-level). All URLNet variants outperform baselines.}
			\label{fig:auc}
		\end{figure}

		The main results of this paper can be seen in Table \ref{tab:main}, and we can further visualize the AUC characteristics in Figure \ref{fig:auc}. In general URLNet based methods significantly outperform the  baseline methods across all metrics (AUC and TPR@FPR). Within the baseline models, we observe that heuristically accounting for sequential information through separate dictionaries in URL Component Tokenization (UCT) is able to improve ther performance over using simple Whole URL BoW Features. Similarly, using other advanced and expert features (PSB and Character Trigrams) is able to give incremental improvement, and the best baseline model is obtained by combining all the features together.

		In contrast, without using any expert or hand-designed features, URLNet methods offer a significant jump in AUC over baselines. It is clear that the proposed URLNet is able to capture several types of semantic and structural information in the URL, which existing methods based on bag-of-words features could not. Within URLNet, we observe the performance of the 3 variants: Character-Level, word-level, and Full. While Character-level and Word-Level URLNet have similar performances, URLNet(Full) largely exploits the positives of both, and provides a more consistently better performance. At low FPRs, word-level URLNet has a better performance than Character-level URLNet, while at higher FPRs, the reverse holds. URLNet(Full) combines the merits of both, and except at FPR=0.0001, it gives a better performance in all other scenarios, including offering a significant boost to the AUC. On the whole, we also observe that increasing training data size from 1 million to 5 million has a positive impact across all metrics.
		
%		\begin{figure}
%			\centering
%			\includegraphics[width=0.5\textwidth]{auc_main}
%			\caption{Area Under ROC Curve characteristics of the proposed URLNet against the baseline models.}
%			\label{fig:auc}
%		\end{figure}
		
		\begin{table*}
			\centering
			\caption{Evaluation of URLNet compared to baselines and special cases of URLNet: Character-level CNN and Word-Level CNNs. All results are on the Test Data of 10 million URLs. The best performances have been bolded.}
			\begin{tabular}{|l|rrrr|r|rrrr|r|}
				\hline
				& \multicolumn{5}{c|}{\textbf{Training Size = 1m}} & \multicolumn{5}{c|}{\textbf{Training Size = 5m}} \\
				\cline{2-11}
				& \multicolumn{4}{c|}{\textbf{TPR @ FPR Level}} &       & \multicolumn{4}{c|}{\textbf{TPR @ FPR Level}} &  \\
				\cline{2-5} \cline{7-10}
				& \textbf{0.0001} & \textbf{0.001} & \textbf{0.01} & \textbf{0.1} & \multicolumn{1}{c|}{\textbf{AUC}} & \textbf{0.0001} & \textbf{0.001} & \textbf{0.01} & \textbf{0.1} & \multicolumn{1}{c|}{\textbf{AUC}} \\
				\hline
				\textbf{Baselines} &       &       &       &       &       &       &       &       &       &  \\
				Whole URL BoW  & 0.1759 & 0.7124 & 0.7915 & 0.8711 & 0.9208 & 0.3189 & 0.7714 & 0.8436 & 0.8995 & 0.9387 \\
				UCT \cite{ma2009identifying}& 0.1431 & 0.7292 & 0.8163 & 0.8890 & 0.9400 & 0.2856 & 0.7890 & 0.8624 & 0.9090 & 0.9548 \\
				PSB \cite{blum2010lexical}& 0.2151 & 0.7560 & 0.8500 & 0.9126 & 0.9549 & 0.1750 & 0.8179 & 0.8834 & 0.9291 & 0.9656 \\
				Character Trigrams\cite{lin2013malicious} & 0.1347 & 0.7326 & 0.8322 & 0.9161 & 0.9611 & 0.2017 & 0.7929 & 0.8682 & 0.9275 & 0.9665 \\
				Combined  & 0.1490 & 0.7499 & 0.8457 & 0.9259 & 0.9669 & 0.2039 & 0.8218 & 0.8883 & 0.9423 & 0.9736 \\
				\hline
				\textbf{URLNet} &       &       &       &       &       &       &       &       &       &  \\
				URLNet (Character-level) & 0.5159 & 0.6876 & 0.8370 & 0.9546 & 0.9828 & 0.6463 & 0.7824 & 0.8991 & 0.9735 & 0.9892 \\
				URLNet (Word-level) & \textbf{0.6337} & 0.7524 & 0.8443 & 0.9297 & 0.9730 & \textbf{0.7312} & 0.8168 & 0.8878 & 0.9595 & 0.9842 \\
				URLNet (Full)  & 0.5849 & \textbf{0.7683} & \textbf{0.8860} & \textbf{0.9722} & \textbf{0.9885} & 0.7160 & \textbf{0.8248} & \textbf{0.9084} & \textbf{0.9858} & \textbf{0.9929} \\
				\hline
			\end{tabular}%
			\label{tab:main}%
		\end{table*}%
		\begin{table*}
		\centering
		\caption{Ablation Analysis: We compare the performance gain from each critical component proposed in URLNet. Both URLNet(Character-level) and URLNet(Word-level) are competitive, and are outperformed by URLNet(Full).}
		\begin{tabular}{|l|rrrr|r|rrrr|r|}
			\hline
			& \multicolumn{5}{c|}{\textbf{Training Size = 1m}} & \multicolumn{5}{c|}{\textbf{Training Size = 5m}} \\
			\hline
			& \multicolumn{4}{c|}{\textbf{TPR @ FPR Level}} &       & \multicolumn{4}{c|}{\textbf{TPR @ FPR Level}} &  \\
			\cline{2-5} \cline{7-10}
			& \textbf{0.0001} & \textbf{0.001} & \textbf{0.01} & \textbf{0.1} & \multicolumn{1}{c|}{\textbf{AUC}} & \textbf{0.0001} & \textbf{0.001} & \textbf{0.01} & \textbf{0.1} & \multicolumn{1}{c|}{\textbf{AUC}} \\
			\hline
			URLNet (Character-level) &&&&&&&&&& \\
			$\hookrightarrow$ Character-level CNN & 0.5159 & 0.6876 & 0.8370 & 0.9546 & 0.9828 & 0.6463 & 0.7824 & 0.8991 & 0.9735 & 0.9892 \\
			\hline
			URLNet (Word-level) &&&&&&&&&& \\
			Word CNN & \textbf{0.6337} & 0.7524 & 0.8443 & 0.9297 & 0.9730 & \textbf{0.7312} & \textbf{0.8168} & 0.8878 & 0.9595 & 0.9842 \\
			$\hookrightarrow +$ Special Characters as Words & 0.5957 & 0.7425 & 0.8496 & 0.9537 & 0.9832 & 0.7179 & 0.8108 & 0.8832 & 0.9649 & 0.9853 \\
			$\quad\quad \hookrightarrow +$ Character-Level Words& 0.6172 & \textbf{0.7694} & \textbf{0.8792} & \textbf{0.9664} & \textbf{0.9865} & 0.6937 & 0.8007 & \textbf{0.8973} & \textbf{0.9673} & \textbf{0.9878} \\
			\hline
			URLNet (Full) &       &       &       &       &       &       &       &       &       &  \\
			$\hookrightarrow$ Character +  Word & \textbf{0.6239} & 0.7565 & 0.8722 & 0.9620 & 0.9864 & \textbf{0.7261} & 0.8169 & 0.8999 & 0.9769 & 0.9907 \\
			$\quad \hookrightarrow + $ Special Character Words & 0.6008 & 0.7512 & 0.8634 & 0.9643 & 0.9853 & 0.6901 & 0.8104 & 0.8996 & 0.9797 & 0.9918 \\
			$\quad \quad \hookrightarrow +$ Character-level Words & 0.5849 & \textbf{0.7683}& \textbf{0.8860} & \textbf{0.9722} & \textbf{0.9885} & 0.7160 & \textbf{0.8248} & \textbf{0.9084} & \textbf{0.9858} & \textbf{0.9929} \\
			\hline
		\end{tabular}%
		\label{tab:abalationURLNet}%
	\end{table*}%
		
	\subsection{Ablation Analysis}
		
		The performance of different components of URLNet, and the incremental performance gain of each component can be seen in Table \ref{tab:abalationURLNet}. Within URLNet (Word-Level), treating special characters as words and using character-level word embedding improve the AUC score. This validates the usage of special characters as words and the benefits of character-level word embedding. Even though this improvement is consistent with varying training data size, a minor discrepancy occurs at the low levels of FPR, where Simple-Word Level URLNet offers the highest TPR. A possible reason for this is that using special characters as words, and using a character-level word embedding make the performance of the word-level CNN slightly resemble that of Character-level CNN, which is why the performance at high FPRs improves, while at low FPRs, it worsens a bit. The trend of word-level URLNet is carried over to URLNet(Full) where we see steady improvement in AUC as we integrate additional features one at a time (special characters as words, and character-level words). On the whole, URLNet(Full) exploits character-level and word-level information and significantly outperforms both Character-based URLNet and Word-based URLNet.

	\subsection{Visualization}
\begin{figure*}[h]
		\centering
		\includegraphics[width=\textwidth]{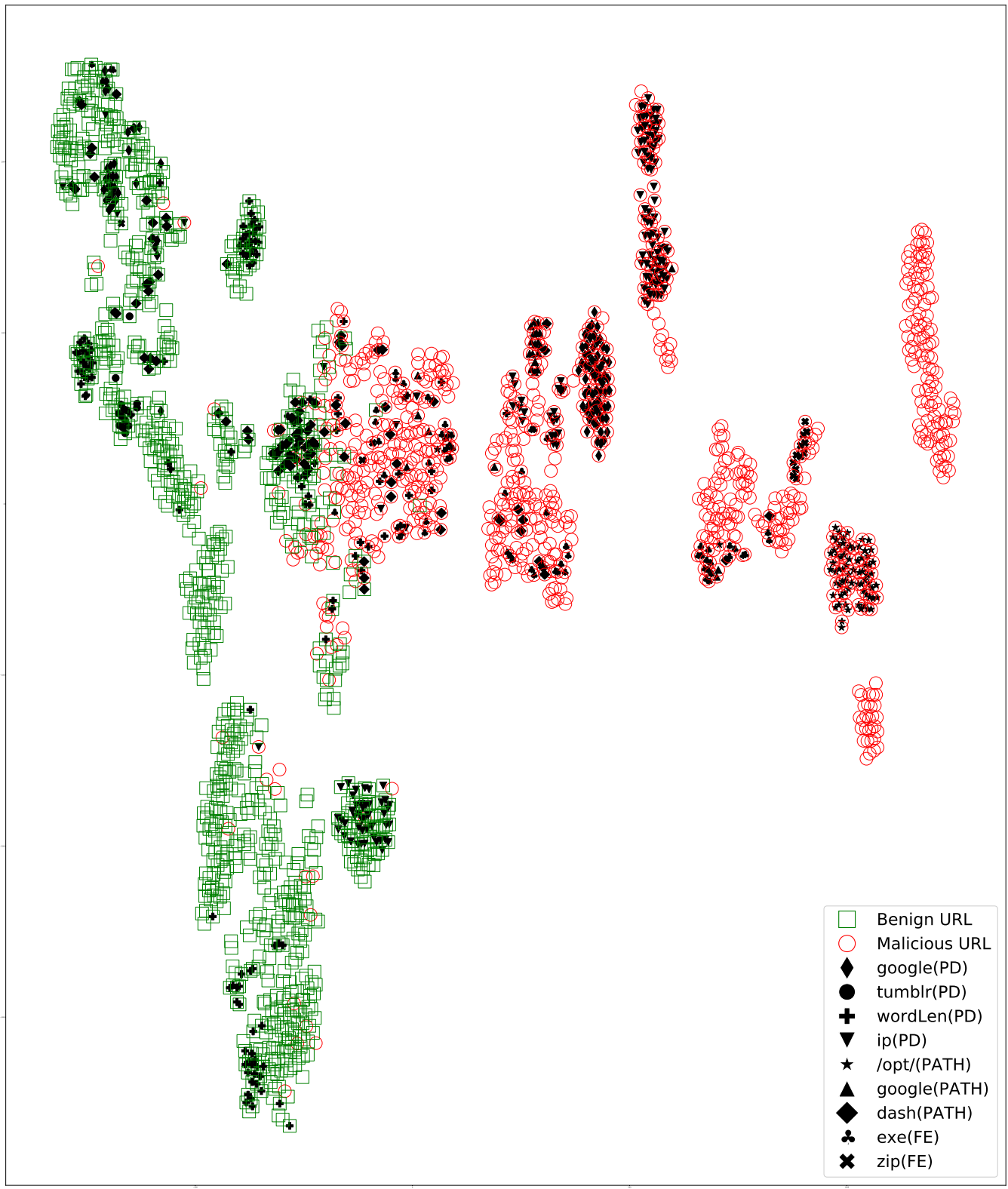}
		\caption{Visualization of feature embedding of sampled URLs using URLNet features. The data points are color-coded by the URL classes: Benign (Green) and Malicious (Red). The markers are also configured to distinguish certain types of lexical patterns in URL components: PD (Primary Domain), PATH (URL Path), and FE (File Extension).}
		\label{fig:urlnetVisual}
	\end{figure*}

\begin{figure*}[h]
		\centering
		\includegraphics[width=\textwidth]{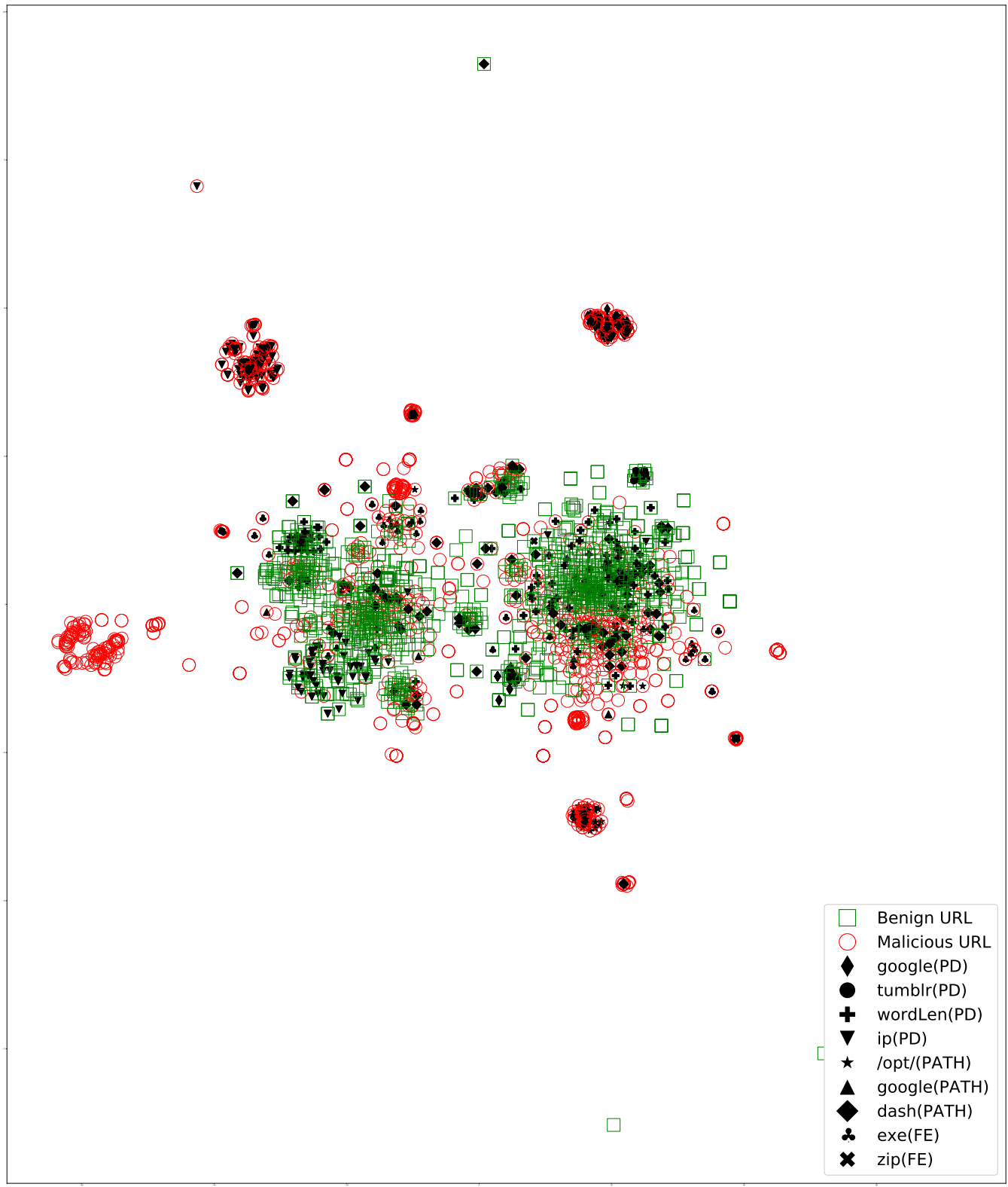}
		\caption{Visualization of feature embedding of sampled URLs using Character Trigrams\cite{lin2013malicious} features. The data points are color-coded by the URL classes: Benign (Green) and Malicious (Red). The markers are also configured to distinguish certain types of lexical patterns in URL components: PD (Primary Domain), PATH (URL Path), and FE (File Extension).}
		\label{fig:baselineVisual}
	\end{figure*}

	In this section, we visualize the embedding features of URLs extracted from the proposed URLNet model and compare with one of the baselines. We randomly sampled 2,000 URLs from the test dataset with balanced distribution of classes (1,000 Benign and 1,000 Malicious) and extracted the feature vectors of these URLs from one of the layers in URLNet (trained on 5 million URLs). Here we selected the feature vectors after concatenation of the outputs of char-level and word-level CNN branches (\emph{Concatenated Char and Word feature vector} layer in Figure \ref{fig:model}), and obtained a 1,024-dimensional vector. For the baseline features, we extract BoW features and expert features being used in the \emph{Character Trigrams}\cite{lin2013malicious}. The baseline feacture vector is 14,604-dimensional. From the extracted feature vectors, we apply t-SNE \cite{maaten2008visualizing} to reduce feature dimension and plot the URLs on a 2-dimensional embedding space. The Figure of the embedded URLs can be seen in Figure \ref{fig:urlnetVisual} and Figure \ref{fig:baselineVisual}.

	As can be seen in Figure \ref{fig:urlnetVisual}, for URLNet, the malicious URLs and benign URLs are clearly seperated into two groups of URLs. Most of Benign URLs are located in the left area of the plot while Malicious URLs are in the right area. Very few data points of Malicious URLs are overlapping with Benign URLs. In the baseline embedding space (Figure \ref{fig:baselineVisual}), the seperation between malicious and benign URLs is not as clear. Many Benign and Malicious URLs are located in the center of the plot and are overlapping with each other. Different from URLNet, in the baseline embedding, many individual data points are scattered around the plot, making it hard to identify which clusters those data points are more likely to belong to. 

	\begin{table*}[]
		\centering
		\caption{Examples of lexical patterns in URLs and example URLs. The lexical patterns are extracted at different parts of the URL string: primary domain, URL path, and file extension.}
		\begin{tabular}{|p{2cm}|l|l|}
		\hline
		\textbf{URL Component}          & \textbf{Lexical Pattern}                & \textbf{Example URL}                                                                                                                                                       \\ \hline
		\multirow{4}{*}{Primary Domain} & contains 'tumblr'                       & http://exampledomain.tumblr.com/                                                                                                                                           \\ \cline{2-3} 
		                                & contains 'google'                       & \begin{tabular}[c]{@{}l@{}}http://www.google.com/urlpath/...\\ http://abcd123googlexyz456.com/urlpath/...\end{tabular} \\ \cline{2-3} 
		                                & contains IP                             & \begin{tabular}[c]{@{}l@{}}http://192.168.0.1/\\ http://192.168.0.1/urlpath/...\end{tabular}                                                                               \\ \cline{2-3} 
		                                & has average word length \textgreater 10 & http://a1ds2dce0b33fdgd425d8fsgg9836c4234d0.exampledomain.net/                                                                                                           \\ \hline
		\multirow{3}{*}{Path}  & contains 'google'                       & \begin{tabular}[c]{@{}l@{}}http://www.exampledomain.com/filename?f=GOOGLEEARTH...\\ http://exampledomain.net/urlpath/googledrive/sub\_dir/...\end{tabular}         \\ \cline{2-3}
		                                & contains '/opt/'                        & http://www.exampledomain.com/opt/...                                                                                                                                       \\ \cline{2-3}
					& contains the \emph{dash} pattern in the last path token                       & \begin{tabular}[c]{@{}l@{}}http://exampledomain.com/urlpath/abc-123-fff-456-... \end{tabular}         \\ \hline
		\multirow{2}{*}{File Extension} & Includes a file with extension 'exe'    & \begin{tabular}[c]{@{}l@{}}http://exampledomain.net/urlpath/filename.exe\end{tabular}                                              \\ \cline{2-3} 
		                                & Includes a file with extension 'zip'    & \begin{tabular}[c]{@{}l@{}}http://exampledomain.com/urlpath/filename.zip\end{tabular}                                              \\ \hline
		\end{tabular}
		\label{tab:lex_pattern}
	\end{table*}

In addition to Benign/Malicious separation, we also observe several clusters appearing in the plots. We further analyze some of these clusters to identify potential patterns in the URL strings which may possibly be indicative of malicious or benign nature of a URL.
We highlight some of the data points with different markers (in black), each of which indicates a type of lexical pattern in the URL string. Refer to Table \ref{tab:lex_pattern} for the details of the lexical patterns and example URLs. Analysis of such patterns could be useful in a deeper understanding of the Malicious URL properties. For example, URLs that contain phrases such as 'tumblr' in URL domain, '/opt/' in the URL path, or '.exe' file extension, are clustered together.
For phrases that appear at different parts of the URL such as 'google', we are still able to distinguish two clusters of URLs, one for 'google' in the URL domain, and one for 'google' in the URL path, as can be seen in Figure \ref{fig:urlnetVisual}, despite not making any distinction between different URL components during training. Similarly, without usage of expert features, URLNet obtains meaningful representation and embeds those URLs where primary domain contains an IP together; or if the length of PD > 10, the URLs get clustered together.

%In particular, for URLs with the \emph{dash} pattern in the last path token (i.e. last path token with alphanumeric phrase alternate with the character '-'), they are clustered together in URLNet better than in the baseline method. This shows the merit of URLNet in using both special characters, such as '-', together with normal words in convolutional neural networks.

Finally, with smaller number of dimensions than the baseline feature vector and without the need to obtain expert features, URLNet feature vector is lightweight and efficient to represent the URL. For data size of 2,000 URLs, URLNet feature vector is about 14 times smaller than the baseline feature vector. Hence, URLNet does not suffer from the memory constraints to process and store the vectors which can be used for further downstream tasks. In contrast to the baselines, the number of dimensions of URLNet feature vector does not vary by the data size, making it easy to obtain and process embedding features of large scale datasets (millions of URLs).  

\section{Related Work}
For a comprehensive review on Malicious URL Detection using Machine Learning, see \cite{sahoo2017malicious}.
Here, we briefly discuss two most related topics for our work: Feature Representation for Malicious URL Detection, and Deep Learning (particularly for Natural Language processing). 

\subsection{Feature Representation for Malicious URL Detection}
Before training a prediction model, a raw URL is typically converted to a suitable feature vector $\u \rightarrow \x$, so that it can be interpreted by traditional machine learning models. This feature representation needs to be chosen carefully, as the classification models work on the premise that distributions of the features for malicious and benign URLs are different. Researchers have proposed several types of features for this task, including blacklist features \cite{ma2009beyond,felegyhazi2010potential}, lexical features \cite{kolari2006svms,ma2009beyond,ma2009identifying, zhao2013cost}, host-based features \cite{mcgrath2008behind,ma2009beyond,chiba2012detecting}, content features \cite{zhang2007cantina,canali2011prophiler,xiang2011cantina+}, and context and popularity based features \cite{choi2011detecting,lee2012warningbird,cao2014detection}. Blacklist features use the presence of a URL in a blacklist as a feature, as they could be strong indicators. Lexical Features focus on string properties of the URL, e.g. length of URL, number of special characters, types of words that appear in the URL string, alpha-numeric distribution of characters, etc. Host-based features are those derived from the host-name properties of the URL including information such as IP Address, WHOIS information, Geographic location, etc. Content features are those that require explicitly visiting and downloading the content hosted by the URL, in order to obtain information such as HTML and JavaScript features. Context and popularity features correspond to information about where the URLs have been shared on social media, or their ranking and popularity scores. Many researchers have used a combination of some of these features, which was often determined through expert domain knowledge.

Obtaining features from URLs can be an expensive task from the perspective of security threats and engineering overhead. For example, obtaining content-based features can be very slow, and at the same time is highly risky. Moreover, identifying which features are useful requires expert domain knowledge. Consequently, usage of information  directly obtainable from the raw URL was popularized \cite{ma2009beyond,ma2009identifying}. It was shown that lexical features, which are the easiest to obtain, gave competitive performances \cite{ma2009beyond,blum2010lexical}. In our work we focused primarily on the lexical features to obtain the feature representation for the URLs. Among lexical features, various types of information can be obtained from the URL. While some statistical properties of the URL string, such as length of the URL, number of dots, etc. \cite{kolari2006svms} have been used, the most popular features were Bag of Words, Term Frequency features or n-gram features \cite{kolari2006svms,ma2009beyond,blum2010lexical}. However, none of these methods effectively capture the sequential properties of the URL string (or substrings). Moreover these methods fail to extract useful information from unseen words in the test URLs. 

There have been other advanced lexical features used, such as Kolmogorv Complexity \cite{pao2012malicious}, obfuscation resistant features \cite{le2011phishdef}, intra-url relatedness \cite{marchal2014phishscore}, etc. However, they required substantial feature engineering or expert knowledge, or they were not scalable to millions of URLs, thus reducing their practical applicability. 	

\subsection{Deep Learning}

Deep Learning or Representation Learning has received increasing interest in recent years owing to their success in several applications \cite{krizhevsky2012imagenet,lecun2015deep,schmidhuber2015deep,goodfellow2016deep,he2015deep}. The core idea is to automatically learn the feature representation from raw or unstructured data, in an end-to-end manner without using any hand designed features. Following this principle, we aim to use Deep Learning for Malicious URL Detection, in order to directly learn representation of the raw URL string, without using any hand designed expert features. 

Since we aim to train Deep Networks over lexical features, a closely related area is Deep Learning for Natural Language Processing (NLP). 
Deep learning methods have found success in many NLP tasks: text classification \cite{kim2014convolutional}, machine translation \cite{cho2014learning}, question answering \cite{xiong2016dynamic}, etc. Recurrent neural networks (e.g. LSTM \cite{hochreiter1997long}) have been widely used due to their ability in capturing sequential information.
However, the problems of exploding and vanishing gradients is magnified for them, making them difficult to train. Recently, Convolutional Neural Networks have become excellent alternatives to LSTMs, in particular showing promising performance for text classification using Word-level CNNs \cite{kim2014convolutional} and Character-level CNNs \cite{zhang2015character}.   

There have been very limited attempts at using Deep Learning for Malicious URL Detection. 
We recently noticed a work parallel to ours \cite{saxe2017expose} that attempted to use Character-level CNNs for this task. However, they ignored several types of structural information that could be captured by words in the URLs. In contrast to their work, (i) we consider both word-level and character-level information; (ii) through extensive analysis we show the importance of word-level information in capturing longer temporal patterns; (iii) we develop novel character-level word embedding for effectively utilizing word-level information - in particular handling the presence of too many unique words, and obtaining embeddings for unseen words at test time; and (iv) we train the whole model in a jointly optimized framework. URLNet comprehensively captures the structural information available in the URL String through both character and word-level information. In fact, \cite{saxe2017expose} is a special case of our proposed URLNet, where only character-level information is considered. Further, we have even shown that URLNet(Full) is consistently better than just a Character-level URLNet in AUC (and TPR at all levels of FPR - particularly at low FPRs).

\section{Conclusion}
	In this paper we proposed URLNet, a CNN based deep neural network for Malicious URL Detection. Existing approaches mostly used Bag of Words like features, and this caused them to suffer from some critical limitations, including inability to detect sequential concepts in a URL string, requiring manual feature engineering, and inability to handle unseen features in test URLs. We proposed Character CNNs and Word CNNs for this task, and jointly optimized the network. Moreover, we proposed advanced word-embedding techniques which are particularly useful to deal with rare words, a problem usually observed in malicious URL Detection tasks (and not in traditional NLP tasks). This approach also allowed URLNet to learn embeddings from unseen words at test time, and exploit subword information. Our approach worked in an end to end manner without requiring any expert features.

\clearpage

\section*{Acknowledgements}
We are very grateful to our collaborators VirusTotal, for providing us access to their data, without which this research would not have been possible. 

\bibliographystyle{ACM-Reference-Format}
\bibliography{mud_bib} 

\end{document}